# Standardization of $^{241}$Am by Digital Coincidence Counting, Liquid Scintillation Counting and Defined Solid Angle counting

C. Balpardo[1], M. E. Capoulat, D. Rodrigues and P. Arenillas

*Laboratorio de Metrología de Radioisótopos, CNEA, Buenos Aires, Argentina*

## Abstract

The nuclide $^{241}$Am decays by alpha emission to $^{237}$Np. Most of the decays (84.6 %) populate the excited level of $^{237}$Np with energy of 59.54 keV. Digital Coincidence Counting was applied to standardize a solution of $^{241}$Am by alpha-gamma coincidence counting with efficiency extrapolation. Electronic discrimination was implemented with a pressurized proportional counter and the results were compared with two other independent techniques: Liquid Scintillation Counting using the logical sum of double coincidences in a TDCR array and Defined Solid Angle Counting taking into account activity inhomogeneity in the active deposit. The results show consistency between the three methods within a limit of a 0.3%. An ampoule of this solution will be sent to the International Reference System (SIR) during 2009. Uncertainties were analysed and compared in detail for the three applied methods.

## Introduction

Among the existing alternatives for the standardization of radioactive sources, the coincidence counting method is one of the most widely used (Campion, 1959; Baerg, 1966) and it has many variants (Bobin, 2007; Keightley and Park; 2007). The Digital Coincidence Counting (DCC) technique used in this work has advantages over the conventional coincidence method, especially in

[1] . Corresponding author. Tel: +54 011 6779 8279; Fax: +54 011 6779 8491
E-mail address: balpardo@cae.cnea.gov.ar (C. Balpardo)

the simplification of the measurement routine.

This paper discusses the application of the DCC method to nuclides decaying by alpha transitions followed by gamma decay, in particular $^{241}Am$, and compares the results with those obtained by defined solid angle counting and Liquid Scintillation Counting using the logical sum of double coincidences in a TDCR array.

**Description of the digital coincidence method**

The basic principle of the implemented DCC technique is the measurement of pulse heights from shaping amplifiers along with time-stamps of the pulses with a time resolution of the order of 10 ns. These pulse heights and times-stamps are stored sequentially in the Static Random Access Memory (SRAM) of the system (Park *et al*, 1998), and transferred to the hard disk of a personal computer upon completion of the measurement process. This task is performed by the TAR (Time and Amplitude Recorder), an electronic NIM module provided by ULS-KOREA.

Selection of gamma gates, the imposition of dead and resolution times, electronic discrimination and the counting corrections using the Cox-Isham formulae (Cox and Isham, 1977) are carried out offline by software.

**Experimental**

The experimental setup consisted of a pressurized proportional counter (PPC) for the detection of α particles sandwiched between two 76 mm x 76 mm NaI(Tl) detectors. The PPC was operated at $3x10^5$ Pascal with P10 gas (Ar (90%) and $CH_4$ (10%)). The signals from both channels were sent to the TAR. The signals are digitized by the ADC of the TAR and stored in the RAM of the module. Finally, the signals are sent to a personal computer in the form of two binary files.

The $^{241}Am$ sources were prepared by depositing one or two drops of the active solution onto the centre of a thin gold-coated VYNS film. A drop of LUDOX® was added in order to reduce self absorption effects.

The nuclide $^{241}$Am decays to $^{237}$Np by α transitions, mainly (84.6%) to the 59.54 keV excited level. The decay scheme of $^{241}$Am is shown in Fig. 1.

For the analysis with the DCC software, the selected γ-gate was 58-60 keV and the imposed resolving times and non-extendable dead times (ICRU report, 1994), in both channels were set to 1.00 (1) µs and 10.00 (1) µs, respectively.

The activity of each source was determined by efficiency extrapolation (Baerg, 1973) with an electronic discrimination method. Figure 2 shows the plot of $N_\gamma . N_\alpha / N_c$ vs. $(1-\varepsilon_\alpha)/\varepsilon_\alpha$, where $N_\alpha$ is the number of registered α events and $\varepsilon_\alpha$ is the efficiency for this nuclide of the proportional counter.

$$\frac{N_\alpha N_\gamma}{N_c} = N_0 [1 + a\frac{(1-\varepsilon_\alpha)}{\varepsilon_\alpha}]$$

The activity concentration value obtained was 297.8(12) kBq.g$^{-1}$.

## Comparison with other standardization methods

*Defined Solid Angle Counting (DSAC)*

Five sources were prepared by drop deposition on five electropolished stainless steel discs. A drop of LUDOX ® was added in order to reduce autoabsorption effects.

The geometrical factor was calculated for each source by taking into account the activity distribution in the active deposit, which was obtained by digitalization of an autoradiograph of the source (Fig. 3). The activity distribution was represented by an array of $N$ = 100 x 100 elements. Since the diameter of the source backing is 20 mm, the area covered by each element of the array was 0.2 x 0.2 mm$^2$. For each element, 100 points were randomly generated within the area covered by it, and the geometrical factor was obtained as the mean of the individual values. Geometrical factors for each point were calculated using the series expansion reported by Jaffey (1954). The overall geometrical factor of the source was obtained as the weighted mean of the geometrical factors of all the elements :

$$G = \frac{\sum_i^N I_i \cdot G_i}{\sum_i^N I_i} \quad ; \quad N = 100\,\text{x}\,100$$

where $G_i$ and $I_i$ are the geometrical factor of the i$^{th}$ element and its corresponding intensity respectively. A similar procedure has been used previously (Garcia-Toraño *et al*, 2008).

The activity concentration obtained was 298.2(14) kBq.g$^{-1}$ at the same reference date, and was calculated by averaging the individual values from each source.

All contributions to uncertainty are shown in table 1. For each source, the uncertainty in the geometrical factor was evaluated utilising the source-to-detector distance, the radius of the detector and the position of the active deposit. Uncertainties in the geometrical factors due to relative intensities were neglected. For our particular set of measurements, it was found that in spite of the absolute intensities for each pixel varying between different autoradiographs from the same deposit, the relative intensities - which are the relevant magnitudes for the overall geometrical factor calculation - did not vary appreciably when the measured optical densities were limited to the linear range of the characteristic curve of the film. Moreover, the overall geometrical factor calculated from a set of autoradiographs for a given deposit did not vary appreciably.

*Liquid Scintillation Counting (LSC)*

Three sources were prepared by dissolving the solute in 10 ml of Ultima Gold AB into glass vials of 22 ml. Each source was measured for a duration of 30 minutes in the TDCR system (Arenillas and Cassette, 2006) under the optimal efficiency conditions (2600 Volt PMT tension and 620 Volt focussing tension).

Usually, a program is necessary to calculate the efficiency, but in this case the value reached by the ratio of triple to double coincidences was 0.9912, therefore the efficiency value could be considered approximately equal to unity. Hence, the activity was directly calculated as the ratio between double counting and the measurement time.

Two major uncertainty components were identified, namely: the mass and the double counting statistic contributions. For further details see Table 1.

The resulting activity concentration was 297.3(10) $kBq.g^{-1}$ at the same reference date.

## Discussion and Conclusion

From Table 1, it can be seen that the uncertainty values for the DCC method are compatible with the other methods. For the DCC and TDCR methods, the uncertainty due to the mass determination is dominant, but in DSAC the counting statistics could still be improved.

Figure 4 shows the consistency between the three methods within a limit of 0.3 %. Moreover, all the results are included within the region corresponding to a coverage factor $k = 1$.

In conclusion, the DCC method has been applied to the standardization of $^{241}Am$ successfully. It confirms that this method can be extended to standardize alpha emitters.

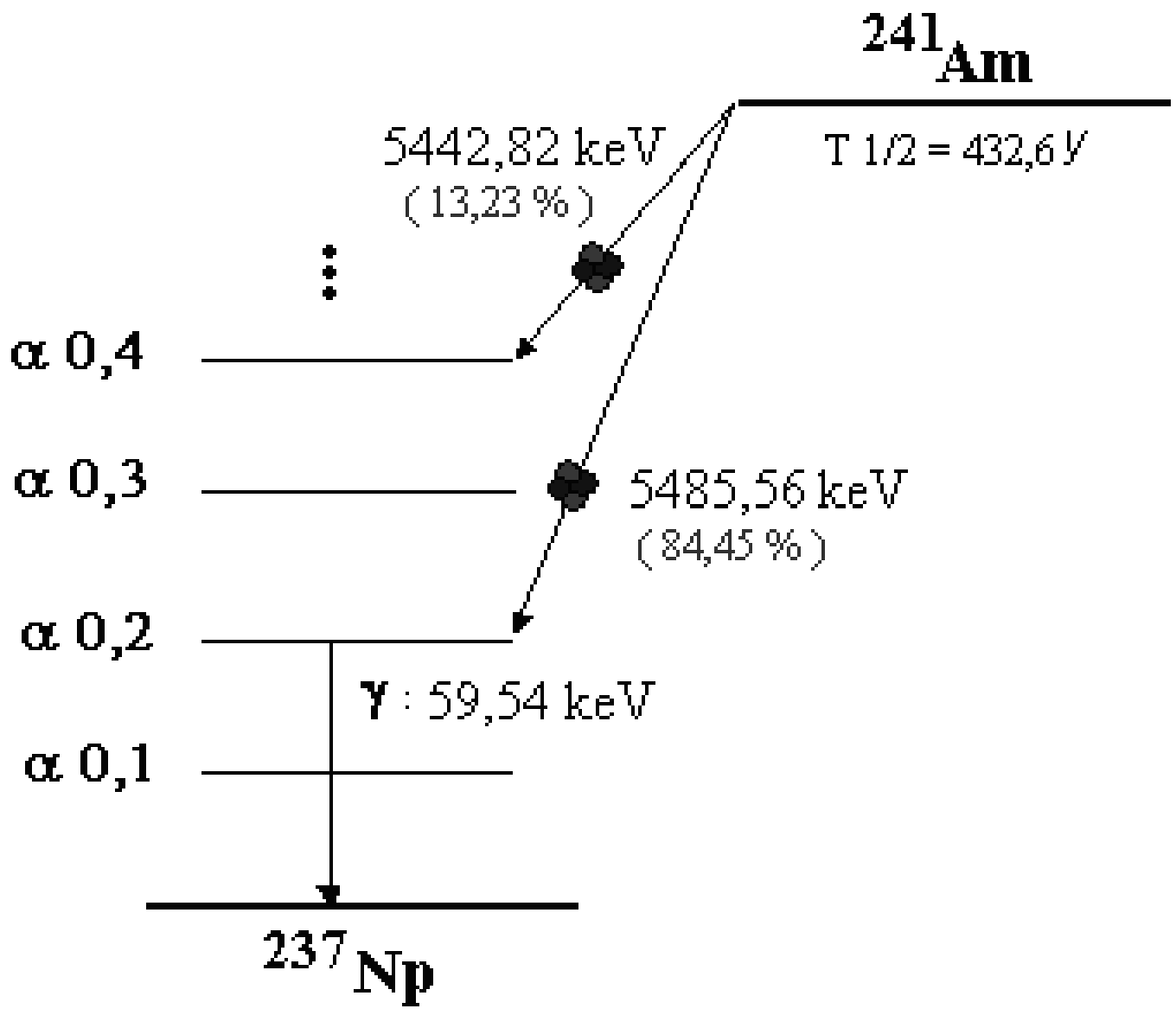

Figure 1: Simplified decay scheme of $^{241}$Am. Only the most intense alpha and gamma emissions are shown.

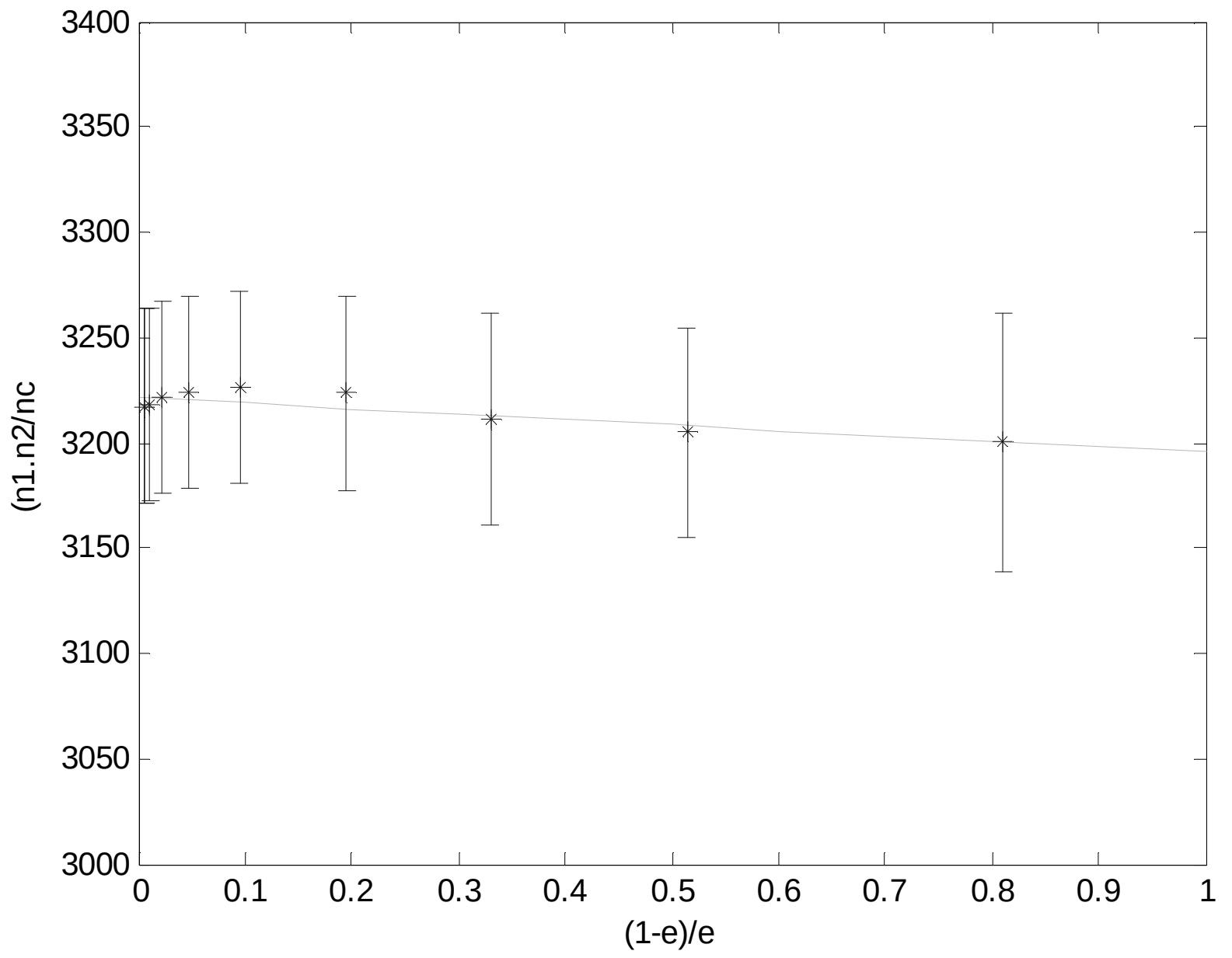

Figure 2. Plot of $\frac{N_\alpha . N_\gamma}{N_c}$ versus $\frac{1-\varepsilon_\alpha}{\varepsilon_\alpha}$ obtained by DCC. The full line shows a linear fit to the data.

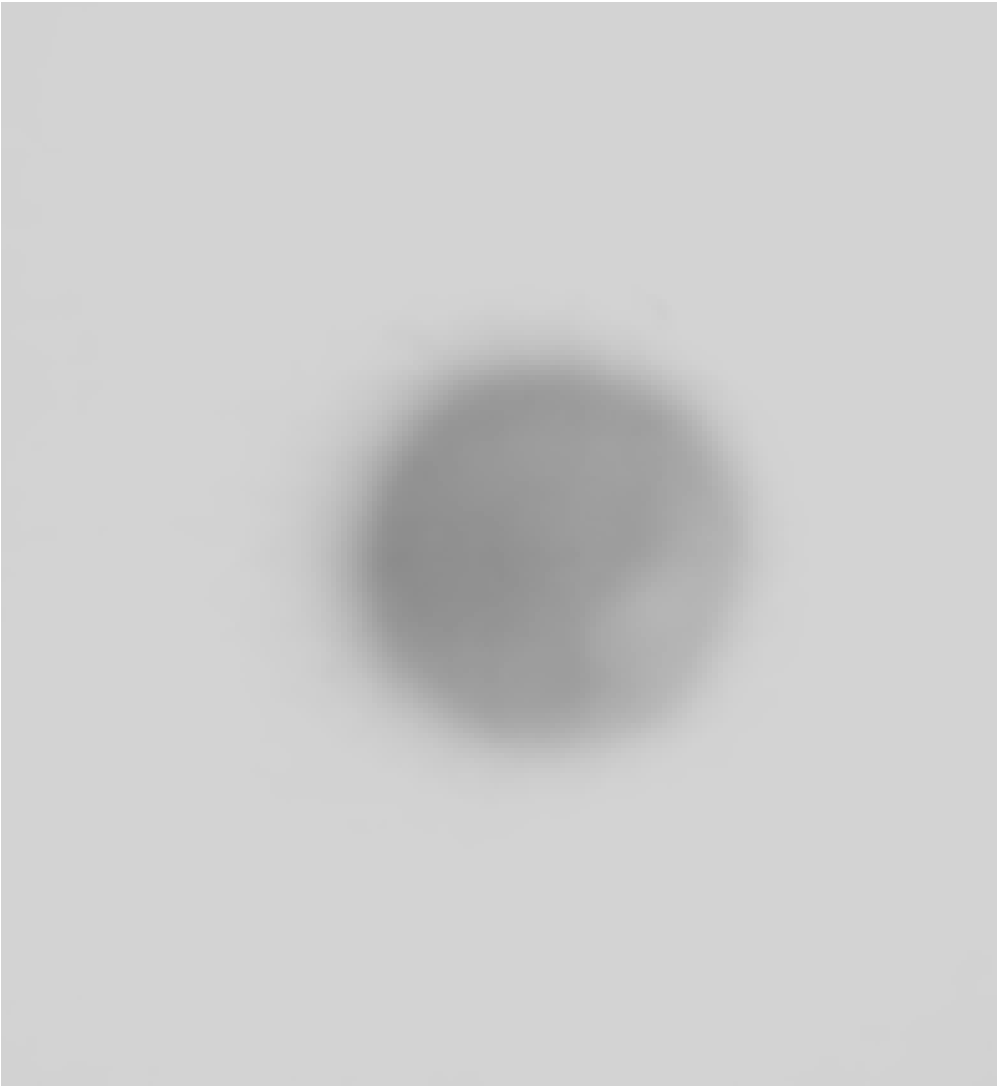

Figure 3.a.

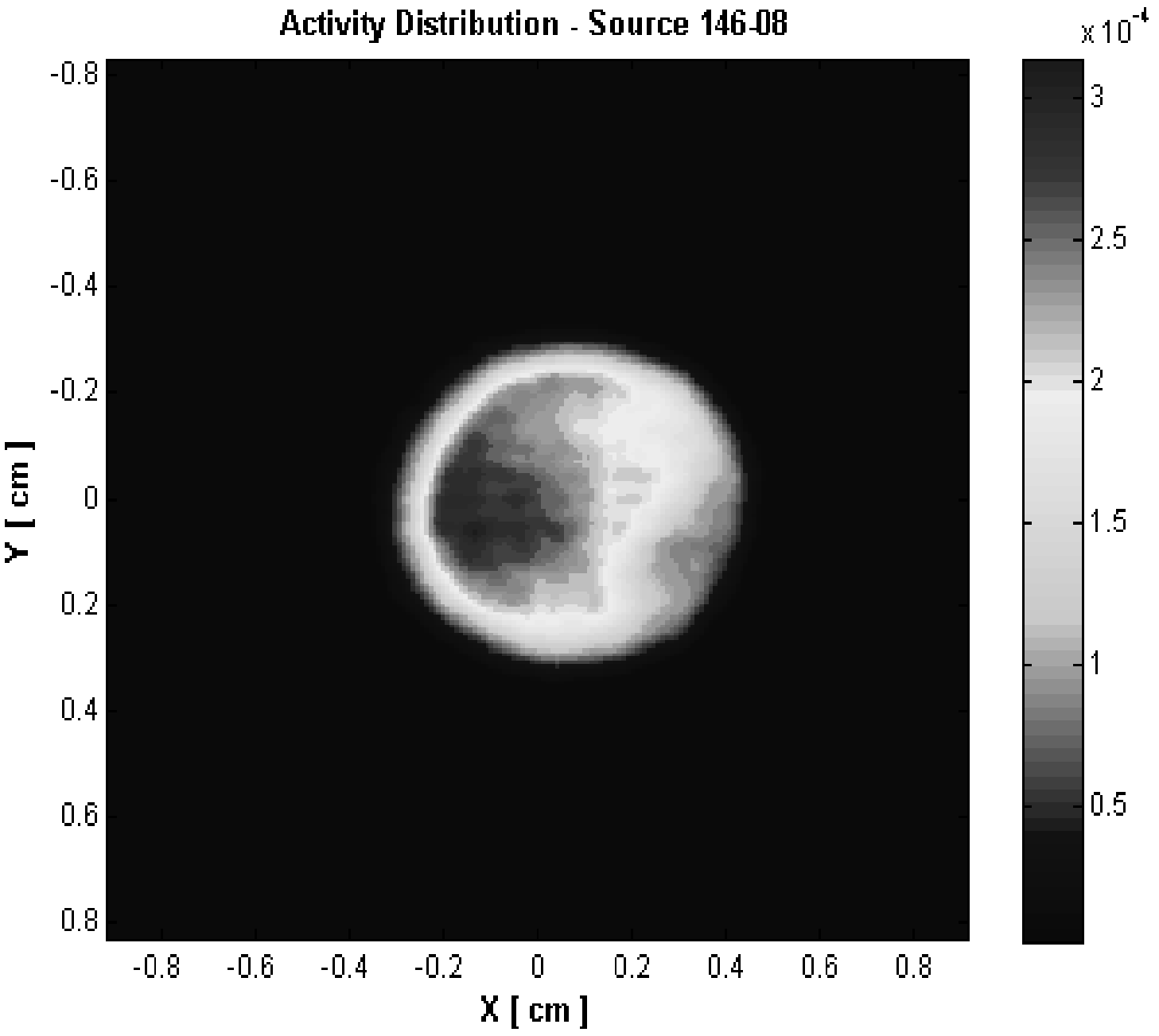

Figure 3.b

Figure 3: Autoradiography of one of the sources used to standardize $^{241}$Am by DSAC

(a). Relative-to-total activity values for each point within the active deposit of the same source

(b). The diameters of the source backing and the active deposit are 2 cm and about 0.7 cm Respectively

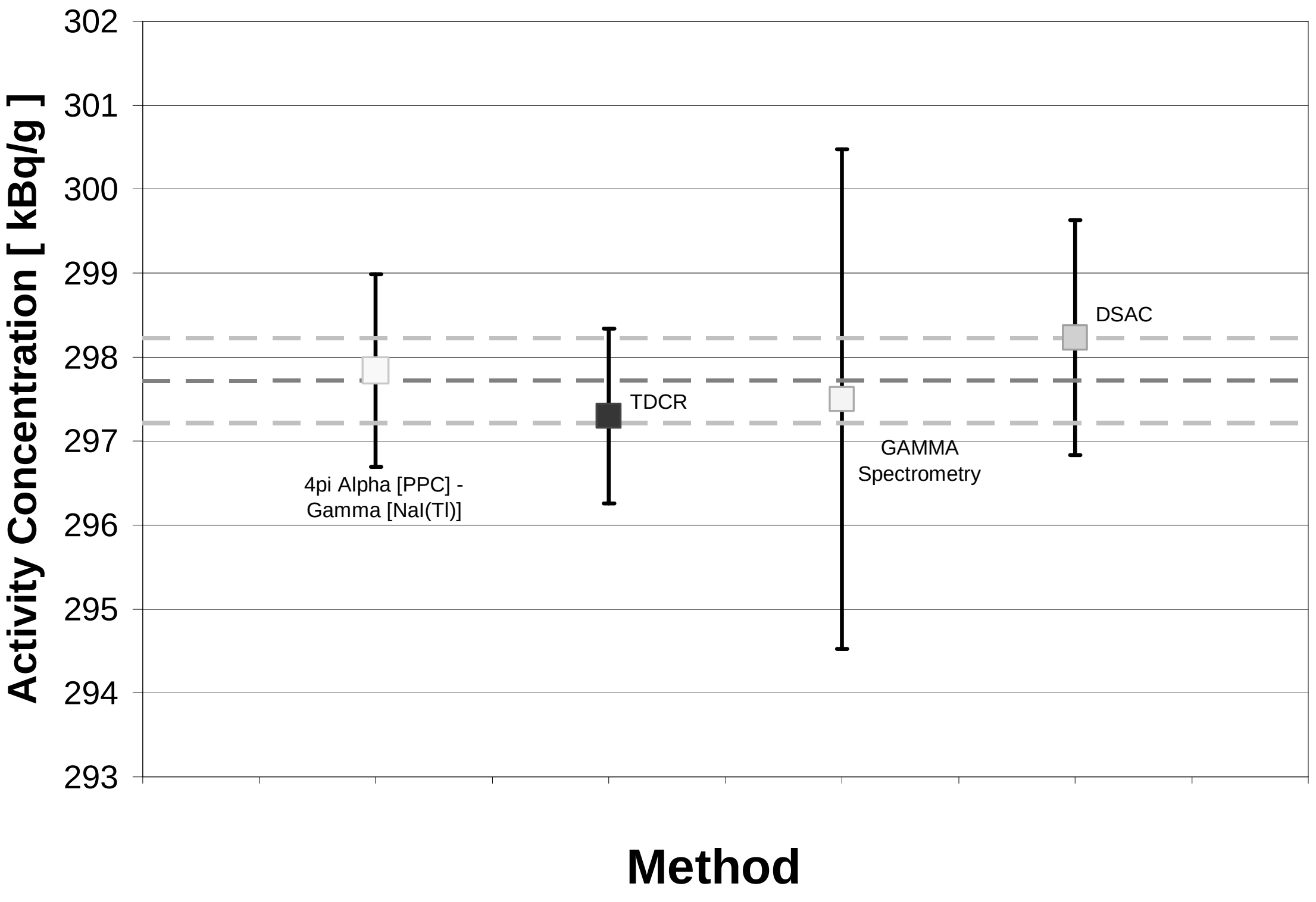

Figure 4: Comparison of the activity concentrations obtained by DCC, LSC and DSAC. For completeness, the activity concentration measured by gamma spectrometry is included. All values are within a 0,3 % region. The error bars plotted for all methods correspond to *k*=1.

| Source of uncertainty | DCC | DSAC | LSC |
|---|---|---|---|
| Extrapolation | 0.060 | - | - |
| Counting statistics | 0.012 | 0.410 | 0.073 |
| Dead time | <<1e-3 | <<1e-3 | - |
| Resolution time | <<1e-3 | - | - |
| Delay time | <<1e-3 | - | - |
| Background | <<1e-3 | 0.007 | - |
| Mass | 0.387 | 0.230 | 0.345 |
| Self absorption | - | <<1e-3 | - |
| Geometrical factor | - | 0.087 | - |
| Backscattering | - | 0.002 | - |
| **Combined uncertainty [%]** | **0.39** | **0.48** | **0.35** |

Table 1: Uncertainty budget for the DCC, DSAC and LSC methods. For DCC and LSC the main component is the mass uncertainty. This contribution could be drastically reduced by using a ultra micro precision weighting balance. Concerning DSAC, the counting statistics contribution is higher than mass uncertainty.